\documentstyle[pre,aps,epsf,multicol]{revtex}

\newcommand{\bleq}{\ifpreprintsty
                   \else
                   \end{multicols}\vspace*{-3.5ex}{\tiny
                   \noindent\begin{tabular}[t]{c|}
                   \parbox{0.493\hsize}{~} \\ \hline \end{tabular}}
                   \fi}
\newcommand{\eleq}{\ifpreprintsty
                   \else
                   {\tiny\hspace*{\fill}\begin{tabular}[t]{|c}\hline
                    \parbox{0.49\hsize}{~} \\
                    \end{tabular}}\vspace*{-2.5ex}\begin{multicols}{2}
                    \fi}
\newcommand{\bcols}{\ifpreprintsty\else\begin{multicols}{2}\fi}
\newcommand{\ecols}{\ifpreprintsty\else\end{multicols}\fi}

\begin{document}
\draft
\title{Fluctuating ``order parameter'' for a quantum chaotic system \\
with partially broken time-reversal symmetry}
\author{S. A. van Langen, P. W. Brouwer, and C. W. J. Beenakker}
\address{Instituut-Lorentz, University of Leiden, P.O. Box 9506, 2300 RA
Leiden, The Netherlands}

\maketitle

\begin{abstract}
The functional $\rho =|\int d\vec{r}\,\psi^2|^2$ of a chaotic wave function
$\psi (\vec{r})$ plays the role of an ``order parameter'' for the
transition between
Hamiltonian ensembles with orthogonal and unitary symmetry. Upon breaking
time-reversal symmetry, $\rho$ crosses over from one to zero. We compute the
distribution of $\rho$ in the crossover regime and find that it has large
fluctuations around the ensemble average. These fluctuations imply long-range
spatial correlations in $\psi$ and non-Gaussian perturbations of eigenvalues,
in precise agreement with results by Fal'ko and Efetov and by
Taniguchi, Hashimoto, Simons, and Altshuler. As a third implication of the
order-parameter fluctuations we find correlations in the response
of an eigenvalue to independent perturbations of the system.
\bigskip
\pacs{PACS numbers: 05.45.+b, 24.60.Ky, 42.25.-p, 73.20.Dx.}
\end{abstract}

\bcols

Wave functions of billiards with a chaotic classical dynamics have been
measured both for classical \cite{stein,prigodin1} and quantum mechanical
waves \cite{chang,folk}.
The experiments are consistent with a $\chi_\beta^2$ distribution of the
squared modulus $|\psi(\vec{r})|^2$ of a wave function at point $\vec{r}$,
the index $\beta=1$ or $2$ depending on whether time-reversal symmetry is
present or completely broken. These two symmetry classes are the
orthogonal and unitary ensembles of random-matrix theory \cite{mehta}.
For a complete description of the experiments one also needs to know
what spatial correlations exist between $|\psi(\vec{r}_1)|^2$ and
$|\psi(\vec{r}_2)|^2$ at two different points and how these correlations
are affected by breaking of time-reversal symmetry.
In the orthogonal and unitary ensembles it is known that
the correlations decay to zero if the distance $|\vec{r}_2-\vec{r}_1|$
greatly exceeds the wavelength $\lambda$ \cite{prigodin2}.

Recently, Fal'ko and Efetov \cite{falko2} managed to compute the
two-point distribution $P_2(p_1,p_2)$ in the crossover regime between
the orthogonal and unitary ensembles. (We abbreviate
$p_i\equiv V|\psi(\vec{r}_i)|^2$, with
$V$ the volume of the system.) They found that the two-point distribution
does not factorize into one-point distributions,
$P_2(p_1,p_2)\neq P_1(p_1)P_1(p_2)$, even if
$|\vec{r}_2-\vec{r}_1|\gg\lambda$. The existence of long-range correlations
in a chaotic wave function came as a surprise.

Two years earlier, in an apparently unrelated paper, Taniguchi, Hashimoto,
Simons, and Altshuler \cite{taniguchi} had studied the response of an
energy level $E(X)$ to a small perturbation of the Hamiltonian
(parameterized by the variable $X$). They discovered a non-Gaussian
distribution of the level ``velocity'' $dE/dX$ in the orthogonal to
unitary crossover. This was remarkable, since the distribution
is Gaussian in the orthogonal and unitary ensembles.

It is the purpose of the present paper to show that these two crossover
effects are two different manifestations of one fundamental phenomenon,
which we identify as {\it order parameter fluctuations}. The ``order
parameter'' is the real number $\rho=|\int d\vec{r}\,\psi^2|^2$ in the interval
$[0,1]$, which equals $1$ ($0$) in the orthogonal (unitary) ensemble.
The possibility of fluctuations in $\rho$ was first noticed by
French, Kota, Pandey, and Tomsovic \cite{french}, but the distribution
$P(\rho)$ was not known.
We have computed $P(\rho)$ in the crossover regime, building on work
by Sommers and Iida \cite{sommers}, and find a broad distribution.
Previous theories for the crossover by {\.Z}yczkowski and Lenz
\cite{zyczkowski}, by Kogan and Kaveh \cite{kogan}, and most recently by
Kanzieper and Freilikher \cite{kanzieper} amount to a
neglect of fluctuations in $\rho$, and thus imply the absence of
long-range correlations in $\psi (\vec{r})$
and a Gaussian distribution of $dE/dX$. Conversely, once the
fluctuations of the order parameter are properly accounted for, we
recover the distant correlations and non-Gaussian distribution of
Refs.\ \cite{falko2,taniguchi}, and find a novel correlation between
level velocities for independent perturbations of the Hamiltonian.

We start from the Pandey-Mehta Hamiltonian \cite{mehta,pandey} for a
system with partially broken time-reversal symmetry,
\begin{equation}
\label{hamiltonian}
H = S + i\alpha\left(2N\right)^{-1/2}A,
\end{equation}
where $\alpha$ is a positive number, and $S$ ($A$) is a symmetric
(anti-symmetric) real $N\times N$ matrix. The matrix $S$ has the
Gaussian distribution
\begin{equation}
P(S)\propto \exp\left(-\case{1}{4}Nc^{-2}{\rm Tr}\, SS^\dagger\right),
\end{equation}
and the distribution of $A$ is the same.
The real parameter $c$ determines the mean level spacing $\Delta$
at the center of the spectrum for $N\gg 1$, by $c=N\Delta/\pi$.
The distribution of $H$ crosses over from the orthogonal to the unitary
ensemble at $\alpha \simeq 1$.
The wave function $\psi_k$ of the $k$-th energy level at widely separated
points ($|\vec{r}_i-\vec{r}_j|\gg\lambda$) is represented by the
unitary matrix $U$ that diagonalizes $H$:
\begin{equation}
V^{1/2}\psi_k(\vec{r}_i) \rightarrow N^{1/2}U_{ik}.
\end{equation}

Consider now an eigenvector $|u\rangle=(U_{1k},U_{2k},\ldots,U_{Nk})$.
(Since we deal with a single eigenstate, we suppress the level index $k$.)
Following Ref.\ \cite{french} we decompose
$|u\rangle$ in the form
\begin{equation}
\label{decomposition}
|u\rangle ={\rm e}^{i\phi} \left(t|R\rangle+i\sqrt{1-t^2}|I\rangle \right),
\end{equation}
where $|R\rangle$ and $|I\rangle$ are real orthonormal $N$-component vectors,
and $\phi\in [0,\pi/2)$ and $t\in[0,1]$ are real numbers. This decomposition
exists for any normalized vector $|u\rangle$ and is unique for $t\neq 0,1$.
The order parameter $\rho$ is related to the parameter $t$ by
\begin{equation}
\rho =\left|\int d\vec{r}\,\psi_k^2\right|^2\rightarrow\left|\sum_i
U_{ik}^2\right|^2=(2t^2-1)^2.
\end{equation}
In the orthogonal ensemble $t=0$ or $1$, hence $\rho=1$, while in the
unitary ensemble $t=\sqrt{1/2}$ hence $\rho=0$.
In the crossover between these
two ensembles the parameter $\rho$ does not take on a single value but
fluctuates.

\begin{figure}
\epsfxsize=0.9\hsize
\vspace*{-6ex}\epsffile{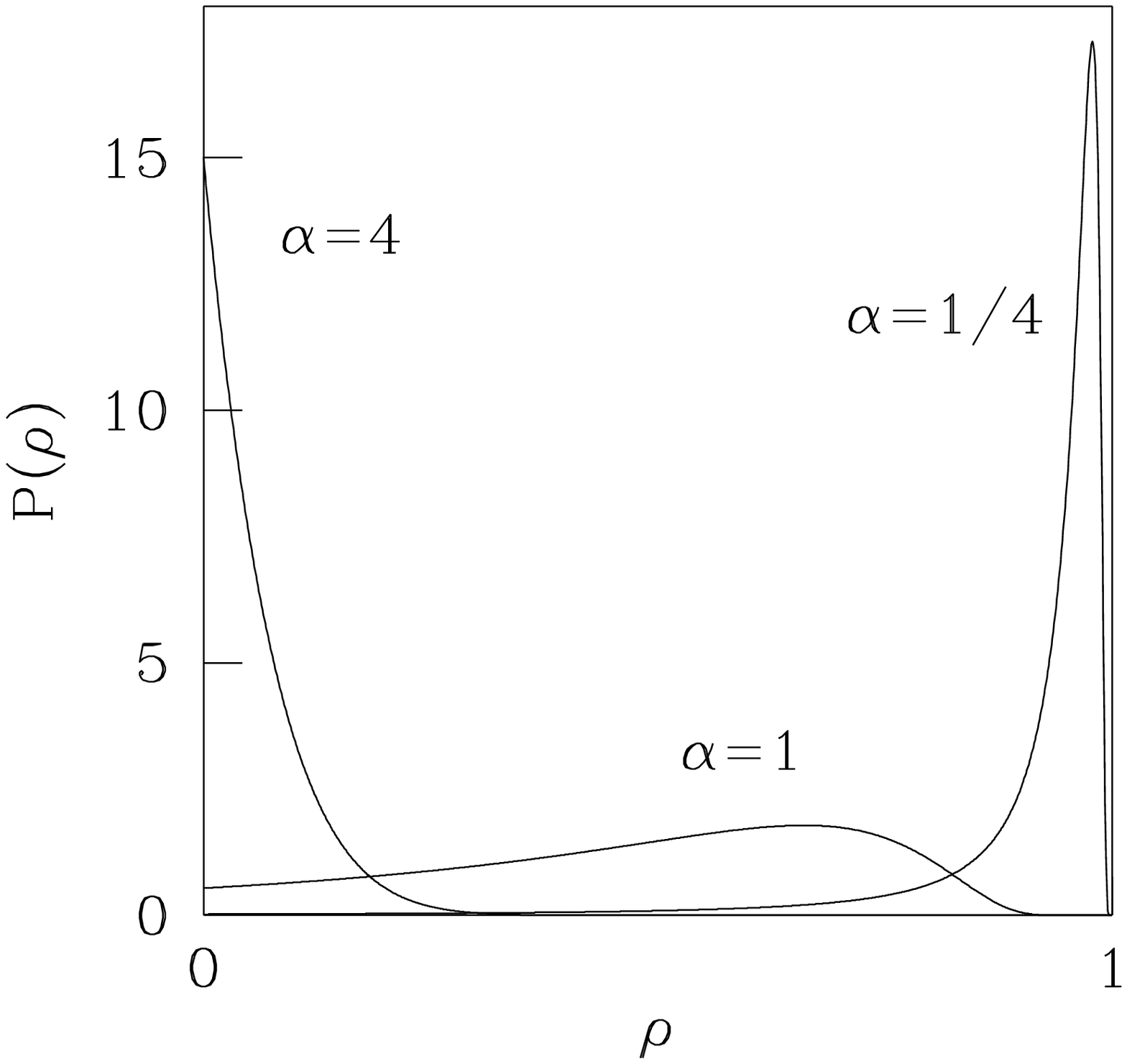}\vspace*{0ex}
\refstepcounter{figure}
\label{fig1}
FIG. \ref{fig1}.
Distribution of
the order parameter $\rho$ for $\alpha=1/4$, $1$, and $4$, computed from
Eq.\ (\ref{result2}).
The crossover from the orthogonal to unitary ensemble occurs when
$\alpha\approx 1$, and is associated with large fluctuations in $\rho$
around its ensemble average.
\end{figure}

To compute the distribution $P(\rho)$ we use a result of Sommers and
Iida \cite{sommers}, for the joint probability distribution of an
eigenvalue $E$ and the corresponding eigenvector $|u\rangle$ of the
Hamiltonian (\ref{hamiltonian}). Substitution of the decomposition
(\ref{decomposition}), and inclusion of the Jacobian for the change
of variables from $|u\rangle$ to $\rho$, gives the expression
\bleq
\begin{mathletters}
\label{intermediate}
\begin{eqnarray}
&& P(\rho)\propto
\frac{(1-\rho )^{N/2-3/2}}
{D^{N/2-1}\sqrt\Lambda}
\left.\left[ \frac{c^2}{N\Lambda} +
\rho\left(\frac{2b_-}{D}\right)^2 \frac{\partial}{\partial b_-}
+\left(\frac{2b_+}{D}\right)^2 \left(\frac{1}{2}\frac{\partial^2}{\partial E^2}
+ \frac{\partial}{\partial b_+} \right)\right]
 Z_{N-2}(E)\right|_{E=0},\\
&&b_\pm=\frac{c^2}{N} \left(1\pm\frac{\alpha^2}{2N}\right),~~
D=4+\frac{2N}{\alpha^2}(1-\rho )\left(1-\frac{\alpha^2}{2N}\right)^2,~~
\Lambda=2+(1-\rho)\left(\frac{2N}{\alpha^2}-1\right), \\
&&Z_N(E)=\frac{1}{N!}\left. \left(b_+\frac{\partial}{\partial\omega}\right)^N
\left(1-\omega
b_-/b_+\right)^{-1}(1-\omega)^{-\case{3}{2}} \left(1+\omega\right)^{-\case{1}{2}}
\exp\left(\frac{-\omega E^2}{(1+\omega)b_+}\right)
\right|_{\omega=0}.
\end{eqnarray}
\end{mathletters}%
\eleq
\noindent
We have set $E=0$, corresponding to the center of the spectrum.
We still have to take the limit $N\rightarrow\infty$.
Expansion of $Z_N(0)$ in a series,
\begin{mathletters}
\begin{eqnarray}
&&Z_N(0)=b_+^N\sum_{k=0}^Na_k\left(\frac{b_-}{b_+}\right)^{N-k},\\
&&a_k=\frac{1}{k!}\left. \frac{\partial^k}{\partial\omega^k}
\left(1-\omega\right)^{-\case{3}{2}}
\left(1+\omega\right)^{-\case{1}{2}}\right|_{\omega=0}
\stackrel{k\gg 1}{\longrightarrow} \sqrt{\frac{2k}{\pi}},
\end{eqnarray}
\end{mathletters}%
and replacement of the summation by an integration, yields
\begin{eqnarray}
Z_N(0)&=&\frac{c^{2N}\sqrt{2/\pi}}{\alpha^2N^{N-3/2}}
\left(e^{\alpha^2/2}+\frac{ie^{-\alpha^2/2}\sqrt{\pi}}
{2\alpha}{\rm erf}(i\alpha )\right)
\label{determinant}
\end{eqnarray}
for $N\gg 1$. Here
${\rm erf}(i\alpha)\equiv 2i\pi^{-1/2}\int_0^\alpha e^{y^2}dy$.
The double energy derivative of $Z_N(E)$ is computed similarly,
but turns out to be smaller by a factor $N$ and can thus be neglected.
The derivatives with respect to $b_\pm$ can be found by differentiation
of Eq.\ (\ref{determinant}). Collecting all terms, we find
\begin{eqnarray}
\label{result2}
&&P(\rho)=(1-\rho)^{-2}\exp\left(\frac{\alpha^2}{\rho -1}\right) \nonumber \\
&&\mbox{}\times\left[ \frac{\alpha^2-1+\rho}{1-\rho}
\left(e^{\alpha^2}+\frac{i\pi^{\frac{1}{2}}\!\!}{2\alpha}{\rm
erf}(i\alpha)\right)
 -\frac{i\alpha\pi^{\frac{1}{2}}\!\!}{2}{\rm erf}(i\alpha )\right].\!
\end{eqnarray}
In Fig.\ \ref{fig1} the distribution of $\rho$ is plotted for three values
of the crossover parameter $\alpha$. It is very broad for $\alpha=1$, and
narrows to a delta function at $1$ ($0$) for $\alpha\rightarrow 0$
($\alpha\rightarrow\infty$).

It remains to show that the long-range wave-function correlations and
non-Gaussian level-velocity distributions of
Refs.\ \cite{falko2,taniguchi} follow from the distribution $P(\rho)$
which we have computed. We begin with the wave-function correlations, and
consider the $n$-point distribution function
\begin{eqnarray}
P_n(p_1,p_2,\ldots,p_n)=\left\langle\prod_{i=1}^n
\delta \bigl( p_i-N|U_{ik}|^2\bigr) \right\rangle.
\label{npointdef}
\end{eqnarray}
We substitute the decomposition (\ref{decomposition}) and do the average
in two steps: First over $|R\rangle$ and $|I\rangle$, and then over $t$.
Due to the invariance of $P(H)$ under orthogonal transformations of $H$,
the vectors $|R\rangle$ and $|I\rangle$ can be integrated out immediately.
In the limit $N\rightarrow\infty$, the components of the two vectors are
$2N$ independent real Gaussian variables with zero mean and variance $1/N$.
Doing the Gaussian integrals we find a generalization of results in
Refs.\ \cite{french,zyczkowski} to $n>1$:
\begin{mathletters}
\label{result1}
\begin{eqnarray}
&&P_n(p_1,p_2,\ldots,p_n)=\int_0^1 d\rho \, P(\rho)
\prod_{i=1}^n F(p_i,\rho), \\
&&F(p,\rho)=(1-\rho)^{-\case{1}{2}}\exp \left(\frac{p}{\rho -1}\right)
I_0\left(\frac{p\sqrt{\rho}}{1-\rho}\right).
\end{eqnarray}
\end{mathletters}%
Here $I_0$ is a Bessel function. We see that long-range spatial
correlations exist only if the distribution $P(\rho)$ of $\rho$ has
a finite width. For example, the two-point correlator
$\langle p_1^2p_2^2\rangle-\langle p_1^2\rangle\langle p_2^2 \rangle$
equals the variance of $\rho$. The approximation of
Ref.\ \cite{zyczkowski} (implicit in Refs.\ \cite{kogan,kanzieper})
was to take $\rho$ fixed at each $\alpha$. If $\rho$ is fixed,
$P_n(p_1,\ldots,p_n)\rightarrow P_1(p_1)\cdots P_1(p_n)$ factorizes,
and hence spatial correlations are absent.
If instead we substitute for $P(\rho)$ our result (\ref{result2}),
we recover exactly the results of Fal'ko and Efetov \cite{falko2,falko1}.

We now turn to the level-velocity distributions. We consider
perturbations of the Hamiltonian (\ref{hamiltonian}) by a real
symmetric (anti-symmetric) matrix $S'$ ($A'$),
\begin{equation}
H' = H + x_{\rm o} S' + x_{\rm u} i A'.
\end{equation}
Here $x_{\rm u}$, $x_{\rm o}$ are real infinitesimals, which parameterize,
respectively, a perturbation that breaks or does not break time-reversal
symmetry. The corresponding level velocities
\begin{eqnarray}
&&v_{\rm o}=\frac{\partial E_k}{\partial x_{\rm o}},
{}~~v_{\rm u}=\frac{\partial E_k}{\partial x_{\rm u}},
\end{eqnarray}
have distributions
\begin{mathletters}
\begin{eqnarray}
P(v_{\rm o})&=&\left\langle \delta \bigl( v_{\rm o}-
\sum_{i,j}U_{ik}^{\vphantom{\ast}}U_{jk}^\ast S'_{ji}\bigr)\right\rangle,\\
P(v_{\rm u})&=&\left\langle \delta \bigl( v_{\rm u}-\sum_{i,j}
U_{ik}^{\vphantom{\ast}}U_{jk}^\ast iA'_{ji}\bigr)\right\rangle.
\end{eqnarray}
\end{mathletters}%
We substitute the decomposition (\ref{decomposition}) for the eigenvector
$U_{ik}$ of $H$ and average first over $S'$ and $A'$, assuming a Gaussian
distribution for these perturbation matrices. After averaging over
$S'$ and $A'$, the eigenvector enters only via the parameter $\rho$. One
finds
\begin{mathletters}
\label{vdistres}
\begin{eqnarray}
P(v_{\rm o})&=&\int_0^1 d\rho\, P(\rho) G_{1+\rho}(v_{\rm o}), \\
\label{vodistres}
P(v_{\rm u})&=&\int_0^1 d\rho\, P(\rho) G_{1-\rho}(v_{\rm u}),
\label{vudistres}
\end{eqnarray}
\end{mathletters}%
where $G_{1\pm\rho}$ is a Gaussian distribution with zero mean and variance
$1\pm\rho$. We have normalized the velocities such that
$\overline{v_{\rm o}^2}=\overline{v_{\rm u}^2}=1$ in the unitary ensemble.
Substitution of
Eq.\ (\ref{result2}) for $P(\rho)$ shows that the distribution of $v_{\rm o}$
coincides with the result of Ref.\ \cite{taniguchi}. However, our
$P(v_{\rm u})$ is different. This is because we have chosen
$A$ and $A'$ to be independent random matrices, whereas they are identical in
Ref.\ \cite{taniguchi}. Independent matrices $A$ and $A'$ are
appropriate for a system with a perturbing magnetic field in a random
direction; Identical $A$ and $A'$ correspond to a system in which
only the magnitude but not the direction of the field is varied.
Eq.\ (\ref{vdistres}) demonstrates that $P(v_{\rm o})$ and $P(v_{\rm u})$
are Gaussians
in the orthogonal and unitary ensembles, since then $P(\rho)$ is a delta
function. In the crossover regime the distributions are non-Gaussian,
because of the finite width of $P(\rho)$. 
The relation (\ref{vdistres}) between the distributions of $v$ and $\rho$
for the GOE--GUE transition is reminiscent of a relation obtained by Fyodorov
and Mirlin for the metal--insulator transition \cite{fyodorov}. The role
of the order parameter $\rho$ is then played by the so-called inverse
participation ratio $I=\int d\vec{r}\,|\psi|^4$. In our system
$NI\rightarrow \rho +2$ for $N\rightarrow\infty$. A difference
with Ref.\ \cite{fyodorov} is that our perturbation matrices are drawn
from orthogonally invariant ensembles, whereas their perturbation is
band-diagonal.

As a final example of the importance of the order-parameter fluctuations
in the crossover regime, we consider the response of the system to
two or more independent perturbations,
\begin{equation}
H'=H+\sum_{i=1}^{m} x_{{\rm o}i} S'_{i} + \sum_{j=1}^{n} x_{{\rm u} j}iA'_{j}.
\end{equation}
For example, one may think of the displacement of $m$ different scatterers,
or the application of a localized magnetic field at $n$ different sites.
Proceeding as before, we obtain the joint probability distribution of the
level velocities $v_{{\rm o}i} = \partial E_k/\partial x_{{\rm o}i}$ and
$v_{{\rm u}j} = \partial E_k/\partial x_{{\rm u}j}$,
\begin{eqnarray}
&&P(v_{{\rm o}1},v_{{\rm o}2},\ldots,v_{{\rm o}m},v_{{\rm u}1},
v_{{\rm u}2},\ldots,v_{{\rm u}n}) \nonumber \\ &&
\ \ = \int_0^1 d\rho\, P(\rho)\prod_{i=1}^m G_{1+\rho}(v_{{\rm o}i})
\prod_{j=1}^n
G_{1-\rho}(v_{{\rm u}j}).
\end{eqnarray}
We see that as a result of the finite width of $P(\rho)$, the joint
distribution of level velocities does not factorize into the individual
distributions (\ref{vdistres}), implying that the response of an energy
level to independent perturbations of the Hamiltonian is correlated.

To summarize, we have computed the distribution of an ``order parameter''
for the breaking of time-reversal symmetry in a chaotic system, defined
as the squared modulus of the spatial average of the wave function
squared. Fluctuations of the order parameter from one
wave function to another exist if time-reversal symmetry is
partially broken. We have shown that these fluctuations imply
long-range wave-function correlations and non-Gaussian eigenvalue
perturbations, thereby unifying two previously unrelated
discoveries \cite{falko2,taniguchi}. A novel manifestation of
the order-parameter fluctuations is the existence of level-velocity
correlations for independent perturbations of the system.

{\it Note added:} We have learned that Y. Alhassid, J. N. Hormuzdiar, and
N. D. Whelan have been working on this same problem, with some overlap
of results.

The authors thank Y. Alhassid, K. B. Efetov, V. I. Fal'ko,
and S. Tomsovic for valuable discussions. This research was supported by
the Dutch Science Foundation NWO/FOM.

\ecols


\begin{references}
\bibitem{stein}
J. Stein, H.-J. St{\"o}ckmann, and U. Stoffregen,
Phys.\ Rev.\ Lett.\ {\bf 75},  53 (1995).
\bibitem{prigodin1}
V. N. Prigodin, N. Taniguchi, A. Kudrolli, V. Kidambi, and S. Sridhar,
Phys.\ Rev.\ Lett.\ {\bf 75}, 2392 (1995).
\bibitem{chang}
A. M. Chang, H. U. Baranger, L. N. Pfeiffer, K. W. West, and T. Y. Chang,
Phys.\ Rev.\ Lett.\ {\bf 76}, 1695 (1996).
\bibitem{folk}
J. A. Folk, S. R. Patel, S. F. Godijn, A. G. Huibers, S. M. Cronenwett,
C. M. Marcus, K. Campman, and A. C. Gossard, Phys.\ Rev.\ Lett.\ {\bf 76}, 1699
(1996).
\bibitem{mehta}
M. L. Mehta, {\it Random Matrices} (Academic, New York, 1991).
\bibitem{prigodin2}
V. N. Prigodin and N. Taniguchi, Mod.\ Phys.\ Lett.\ B {\bf 10}, 69 (1996).
\bibitem{falko2}
V. I. Fal'ko and K. B. Efetov, Phys.\ Rev.\ Lett.\ {\bf 77}, 912 (1996).
\bibitem{taniguchi}
N. Taniguchi, A. Hashimoto, B. D. Simons, and B. L. Altshuler,
Europhys. Lett. {\bf 27}, 335 (1994).
\bibitem{french}
J. B. French, V. K. B. Kota, A. Pandey, and S. Tomsovic,
Ann.\ Phys.\ (N. Y.) {\bf 181}, 198 (1988).
\bibitem{sommers}
H.-J. Sommers and S. Iida, Phys.\ Rev.\ E {\bf 49}, R2513 (1994).
\bibitem{zyczkowski}
K. {\.Z}yczkowski and G. Lenz, Z. Phys.\ B {\bf 82}, 299 (1991).
\bibitem{kogan}
E. Kogan and M. Kaveh, Phys.\ Rev.\ B {\bf 51}, 16400 (1995).
\bibitem{kanzieper}
E. Kanzieper and V. Freilikher, preprint cond-mat/9609081.
\bibitem{pandey}
A. Pandey and M. L. Mehta, Commun. Math. Phys. {\bf 87}, 449 (1983).
\bibitem{falko1}
V. I. Fal'ko and K. B. Efetov, Phys.\ Rev.\ B {\bf 50}, 11267 (1994).
\bibitem{fyodorov}
Y. V. Fyodorov and A. D. Mirlin, Phys.\ Rev.\ B {\bf 51}, 13403 (1995).
\end{references}
\end{document}